# Comparing time and frequency domain estimation of neonatal respiratory rate using pressure-sensitive mats

Shermeen Nizami[1], Amente Bekele[1], Mohamed Hozayen[1], Kim Greenwood[2], JoAnn Harrold[3], James R. Green[1]

[1]Systems and Computer Engineering, Carleton University, Ottawa, Canada
[2]Clinical Engineering, [3]Neonatology, Children's Hospital of Eastern Ontario, Ottawa, Canada
shermeen@sce.carleton.ca, jrgreen@sce.carleton.ca

*Abstract*—Pressure-sensitive mats (PSM) have proved to be useful in the estimation of respiratory rates (RR) in adult patients. However, PSM technology has not been extensively applied to derive physiologic parameters in infant and neonatal patients. This research evaluates the applicability of the capacitive XSensor PSM technology to estimate a range of RR in neonatal patient simulator trials conducted under several experimental conditions. PSM data are analyzed in both the time and frequency domain and comparative results are presented. For the frequency-domain approach, in addition to estimating RR, a measure of confidence is also derived from the relative height of peaks in the periodogram. The study demonstrates that frequency domain analysis of mean-shifted PSM data achieves the best possible RR estimation, with zero percent error, as compared to the lowest achievable RMS error of 1.57 percent in the time domain. The frequency domain approach outperforms the time domain analysis whether examining raw data or those preprocessed by normalizing, detrending and median filtering.

*Keywords*— pressure sensitive mat; neonatal intensive care unit; neonate; respiration rate; simulator

## I. Introduction

There is increasing evidence that the respiratory rate (RR) of a person lying on a pressure-sensitive mat (PSM) can be estimated by analyzing the measured contact pressure data. Past research that demonstrates the utility of PSM data for the estimation of RR has been mainly focused on the adult population, such as [1]–[12]. Two systematic reviews of non-invasive respiratory monitoring in clinical care, conducted almost a decade apart [13], [14], were similarly focused on adult studies. The most recent review, in 2015, concludes that such monitoring has the potential for improved early diagnosis of patient deterioration and the reduction of critical events for patients on the general wards [14]. PSM technology is well-suited to long-term patient monitoring both at home and in hospitals due to it being non-invasive, contactless, and unobtrusive. However, when it comes to the infant and neonatal population, there is a scarcity of research within the field.

In 1976, Franks *et al.* pioneered the idea to measure pressure changes in the thorax region during the respiratory cycle in overnight recordings in infants at home who were up to six months old [15]. They recommended that an under-mattress pressure sensor was the most simple and satisfactory amongst five contactless devices tested, especially in cases where there could be considerable parental apprehension. However, the use of PSM has not yet been evaluated on critically ill term and preterm babies within a hospital's neonatal intensive care unit (NICU). Previously, the use of mechanical sensors such as single-point load cells [16], pressure sensors [17], [18], and piezoceramic sensors [19] has been researched for the extraction of neonatal and infantile RR. The key disadvantage with all these methods is the loss of spatial resolution when compared to PSM technology, where an array of pressure transducers provides a time-varying high-density 2D map of contact pressure. Secondly, sensors that attach to the patient's bed are relatively difficult to transfer should the patient needs to be transferred to another bed.

A number of infant breathing monitors that use mechanical sensors are available in the market including wearables such as Snuza Hero, Levana Oma, Mimo and BabySense [20]. However, these systems are neither licensed as medical devices for use in hospitals in North America nor do they provide application program interfaces for secondary analyses of their datasets.

As we begin to explore the use of PSM as a long-term continuous patient monitoring modality in the NICU, there exist a number of challenges. A major difference that exists in using PSM with the neonatal population, as opposed to adults or older children, is the much smaller mass and pressure ranges [16]. It may be necessary to place the PSM above the mattress but below a crib sheet, since output signals below a mattress can be heavily muted and may feature less distinct localized loading thus reducing the signal-to-noise ratio (SNR) [4]. Thereby, adherence to medical device standards and testing for electrical safety become of utmost importance. In addition, electrical, mechanical, and environmental artifacts decrease the ability to estimate the RR accurately. These include power line interference, electrical noise, patient movements, and vibrations from people walking around the bed, possible air currents from heating, ventilation and air conditioning. Physiologic artifacts are another potentially confounding factor in the detection of the RR signal from PSM data. For example, expiratory grunting distorts the breathing signal and produces an unclear noisy expiration wave [21], [22]. Grunting is a compensatory breathing effort made by a neonate to overcome some lung abnormality. Grunting is produced by the expiration of air through partially closed vocal cords, either intermittently or continuously depending on the severity of the lung disease [23]. Clinicians typically hear the grunting sound produced by a patient. Grunting is one of the symptoms of neonatal respiratory distress syndrome (NRD) and is typically recognized by clinicians when listening to a patient breath [24]. Term and preterm infants presenting with NRD are admitted to the NICU and treated with oxygen supply [25], [26]. NRD accounts for significant morbidity and mortality [27].

In this proof-of-concept study, we collect a comprehensive dataset from a capacitive PSM LX100:36.36.02 (XSensor Technology Corp. Calgary, Canada, XSensor.com) to show that it is possible to extract simulated neonatal RR from PSM data. In our previous work, the XSensor PSM technology has shown a superior dynamic response as compared to the metrological properties of resistive Tekscan and optical S4 PSM technologies [28]. This study compares time and frequency domain approaches for RR estimation with results presented in terms of estimation error and confidence.

A breathing pattern is a time series signal dominated by respiratory modulation [29]. To compare RR estimation results in the time and frequency domains, this study categorizes the dataset based on three experimental conditions: (i) breathing patterns (normal vs. grunting); (ii) position of the simulator (supine vs. prone); and (iii) type of mattress (overhead warmer vs. crib). These data are then pre-processed as follows to mitigate signal artifacts: normalizing by subtraction of average pressure measured across the analysis window width to remove DC components extraneous to the breathing pattern; detrending to remove low-frequency metrological drift; and median filtering for recovery of a smooth breathing pattern. Each stage of pre-processing aims at improving the signal-to-noise ratio (SNR) of the breathing pattern, which is otherwise obscured by noise in the PSM data. The time domain analysis does not require knowledge of the expected RR range. RR estimation in the time domain is based on calculating the number of times the breathing pattern crosses a given threshold. In addition to the time domain, this study analyzes PSM data in the frequency domain to estimate RR. The breathing pattern is analyzed in the frequency domain by fast Fourier transformation (FFT) and subsequent identification of the frequency component that is contributing the largest signal power.

There is value in having independent estimates of RR from multiple patient monitoring modalities, including PSM, for greater data integrity. However, the estimate produced by PSM data may be confounded by several factors that could be electrical, mechanical, environmental and physiologic in nature. This research focuses on the impact of physiologic grunting on the RR estimate by analyzing PSM data acquired during bench testing with neonatal patient simulators. This bench testing constitutes phase one of a larger project that will assess the applicability of PSM technology to patient monitoring in the NICU. Phase two shall include software and systems development guided by real patient data acquired from the NICU. The following section outlines the data collection and acquisition methods. Section III presents results and data analysis. Section IV discusses the research findings followed by conclusions with directions for future work.

## II. METHODS

The bench testing was conducted at the Children's Hospital of Eastern Ontario (CHEO), Ottawa, Canada. The bench testing equipment included a neonatal patient simulator "SimNewB" (Laerdal Medical Canada, Ltd., Toronto, Canada), a Giraffe overhead warmer neonatal bed (GE Healthcare, USA), an open crib, and a capacitive PSM LX100:36.36.02 (XSensor Technology Corp. Calgary, Canada, XSensor.com). The bench testing was conducted by placing "SimNewB" on two different mattresses. One mattress comes with the Giraffe overhead warmer in the size of 65x48x4 cm (25.5L x 19W x 1.5D in), and a crib mattress that is approximately double the size and depth, and is significantly firmer. Fig. 1 shows "SimNewB" lying in the supine position on the PSM in the crib. The "SimNewB" simulator represents a neonate weighing approximately 2790g (6.2 lbs) with a length of 51 cm (21 in). The PSM sensor was placed on top of the bed mattress and covered with a sheet that is normally used in the NICU. The PSM sensor has a density of 1 sensel/0.5 in$^2$ with an overall sensing area of 18 x 18 in$^2$. The PSM connects to an X3 Pro Sensor Pack that feeds into an X3 Pro Electronic Platform that is connected via USB to a laptop running the X3 Pro software. The X3 Pro software was used to record PSM data and video simultaneously. Fig. 1 also shows the contact pressure image produced by the X3 Pro software in one frame during the acquisition of a supine dataset in the crib. The labels indicate the body parts of the simulator on the PSM. The shaded thorax area marks the region of interest for which the data were analyzed.

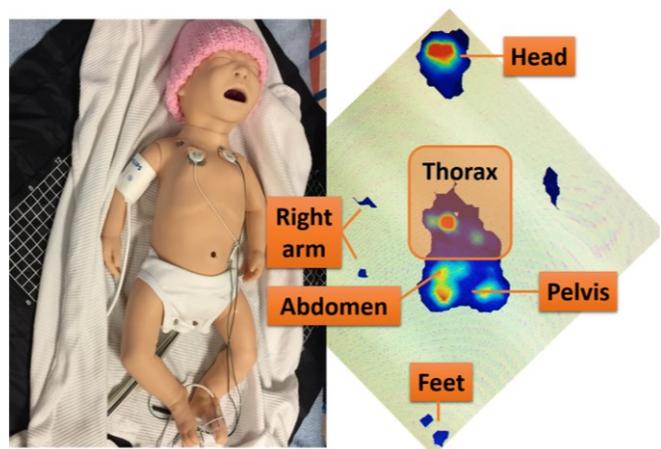

**Fig. 1: SimNewB lying supine on XSensor PSM over a crib mattress, with the pressure image shown on the right**

*A. Data Acquisition and Validation*

A total of twenty trials were conducted, of which, 12 were on an overhead warmer while 8 were measured on the crib mattress, 12 represented normal breathing patterns while 8 represented breathing with grunting, and in 10 experiments, the simulator was lying in the supine position with the remaining 10 in the prone position. The trials were 30-80 seconds long and contact pressure data were acquired at a sampling rate of 20 frames/sec. Breathing is a mechanical function of the simulator, where air from an external compressor is used to cyclically inflate an air sac simulating both lungs. The simulator was set to breathe normally, or with grunting, at three different RR of 45, 60 and 75 breaths per minute (bpm). These RR fall within the ranges observed in neonates, whether preterm or term born, as specified in [22], [30]. Average contact pressure data acquired from the thorax region were analyzed to extract breathing patterns.

The X3 Pro calibrates the lowest noise floor (NF) based on the pressure values induced by the load placed on the mat. Based on the neonatal simulator load the NF is 0.0773 psi in these experiments. It implies that pressure values below 0.0773 psi are excluded from the average pressure calculations produced by the PSM technology. As shown in Table I, the RR estimation results in the time and frequency domains are compared across three experimental conditions with the following characteristics: (i) normal breathing patterns versus breathing patterns that include grunting; (ii) whether the simulator is lying in a supine or a prone position on the PSM; and (iii) the type of mattress underneath the PSM (overhead warmer or crib mattress).

*B. Signal Pre-processing*

Signal pre-processing was applied across the entire recording length in each trial. The acquired PSM data were pre-processed to suppress signal artifacts and isolate the breathing pattern. Normalization is carried out to remove the DC bias similar to the method in [31]. The DC bias is caused by static forces from the load placed on the PSM. To normalize, the average of all data points in the analysis window is calculated and subtracted from each data point in the window. The DC signal causes a very large peak at zero in the periodogram, thus overshadowing the power of the fundamental frequency of the respiratory cycle. Therefore, it is necessary to normalize the average contact pressure data. The normalized signal is detrended using MATLAB's *detrend* function to remove the slow signal fluctuations due to sensor drift. We have defined and evaluated drift and other metrological properties of PSM in our past research [28]. Detrending removes the low-frequency noise in the frequency domain, hence rendering a cleaner periodogram. Finally, the data are median filtered to smooth the signal, remove higher frequency noise components and recover the breathing pattern. The three stages of signal pre-processing, namely normalization, detrending and median filtering all aim to improve the overall SNR.

*C. Time Domain Analysis*

The time-domain analysis is applied to the raw average pressure signal acquired from the PSM prior to pre-processing, as well as to the normalized, detrended, and the median filtered datasets. The time domain signal is analyzed for the number of times it crosses a set threshold. The threshold is set to the 75th percentile value of the raw data and to half the value of the 75th percentile of the pre-processed time series data. The method used to count the respiratory peaks is similar to that of [32]. The RR is estimated by dividing the number of threshold crossings by twice the number of samples or frames in the analysis window and multiplying it by sixty times the frame rate to get a value in bpm. Finally, the percentage error of the RR estimate in the time domain (RR-TD$_E$) is calculated by comparing it with the RR value set on the simulator. The root mean square (RMS) of RR-TD$_E$ for the raw and pre-processed datasets for various experimental conditions are reported in Table I.

*D. Frequency domain analysis*

For infants with a corrected age in the range of 1 to 79 weeks, the respiratory signal lies in the low-frequency band [30]. In this research the neonatal simulator's RR was set to 45, 60 and 75 bpm, corresponding to frequencies of 0.75-1.25 Hz. Following the methods in [31], [32], the time-series data are filtered using a second-order Butterworth filter with a passband of 0.3 to 1.5 Hz. MATLAB's *fft* function is applied to the bandpass signal from which the fundamental frequency, i.e., the frequency with the highest power contribution, *a*, is selected and then multiplied by 60 to estimate the RR in bpm. The second largest peak, *b*, within the passband is also determined. The percentage error (RR-FD$_E$) is computed identically to RR-TD$_E$. A confidence measure (L$_C$) is computed as the ratio of the highest and the second highest peaks in the periodogram, i.e., L$_C$ = $a/b$. Larger values of L$_C$ indicate greater RR estimation confidence. This frequency-domain analysis is applied to the raw average pressure prior to pre-processing, as well as to the normalized, detrended, and the median filtered datasets. Table I lists the RMS of RR-FD$_E$ and L$_C$.

Table I: Aggregate results of time and frequency domain analyses

| Experimental Condition | Number of experiments | Raw Data | | Normalized Data | | Detrended Data | | Median Filtered Data | |
|---|---|---|---|---|---|---|---|---|---|
| | | RR-FF$_E$ | RR-TD$_E$ | L$_C$ | RR-TD$_E$ | L$_C$ | RR-TD$_E$ | L$_C$ | RR-TD$_E$ |
| **Position** | | | | | | | | | |
| Supine | 10 | 64.31 | 122.82 | 1.69 | 133.42 | 1.69 | 131.98 | 1.69 | 13.23 |
| Prone | 10 | 0.53 | 9.35 | 1.57 | 0.70 | 1.57 | 0.70 | 1.56 | 1.57 |
| **Breathing Pattern** | | | | | | | | | |
| No Grunting | 12 | 39.34 | 59.59 | 1.56 | 68.11 | 1.56 | 64.94 | 1.57 | 4.45 |
| With Grunting | 8 | 45.39 | 108.18 | 1.69 | 114.51 | 1.69 | 113.56 | 1.66 | 12.17 |
| **Mattress Type** | | | | | | | | | |
| Overhead Warmer | 12 | 38.86 | 83.57 | 1.69 | 99.42 | 1.69 | 97.01 | 1.68 | 10.91 |
| Crib | 8 | 1.21 | 86.09 | 1.53 | 80.97 | 1.53 | 80.03 | 1.51 | 4.92 |

## III. RESULTS

### A. Data Acquisition and Validation

The "SimNewB" simulator model's weight and length are representative of those of newborn babies [22][33]. Fig. 1 is a pressure image from one frame of PSM data of the simulator lying in the crib in the supine position. Fig. 2 shows results from a trial in which SimNewB is lying in a supine position on the crib mattress and breathing normally at 45 bpm that corresponds to a fundamental frequency of 0.75 Hz. The breathing cycle is visible in the raw PSM time series data acquired from the thorax region as shown in Fig. 2a.

### B. Signal Pre-processing

Fig. 2 also shows time series plots at each stage of pre-processing. It is interesting to note the cyclical fluctuation in the signal around the normalized mean value of zero in Fig. 2b. These fluctuations illustrate drift, which is a metrological property of the PSM's capacitive technology. Drift is reduced by detrending (Fig. 2c). Fig. 2d shows the smoothing effect of the median filter, resulting in a cleaner breathing pattern.

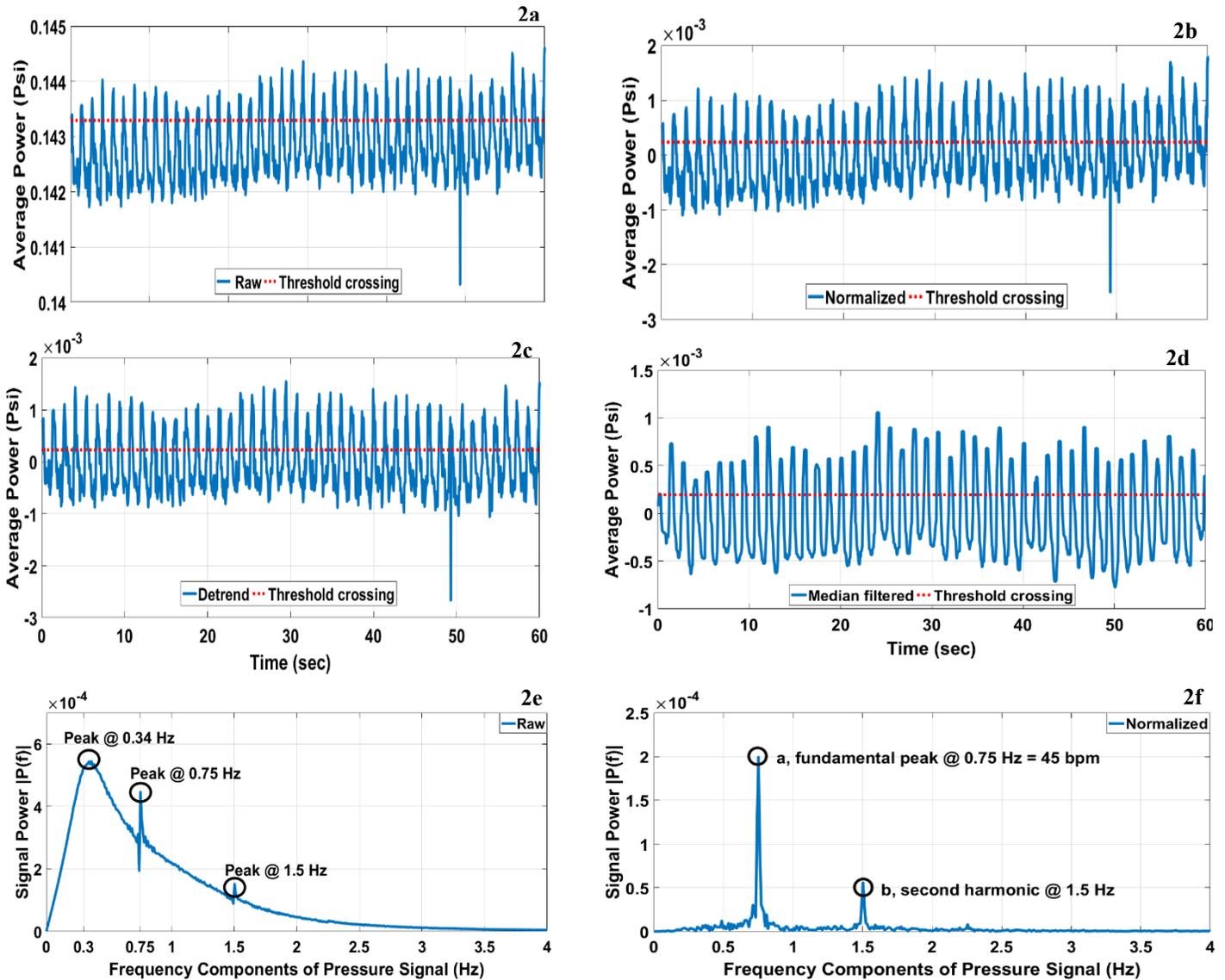

Fig. 2: For SimNewB lying supine on crib, normal RR at 45 bpm, time series of average power (psi) are shown (a) raw, (b) normalized, (c) detrended, and (d) median filtered data; signal power in the frequency domain is shown after applying a bandpass filter to (e) raw and (f) normalized data.

*C. Data Analysis*

In all four time series plots in Fig. 2, there is a horizontal line that represents the threshold crossing that is used to count the number of RR peaks. Fig. 2e-f illustrate the frequency domain representations showing the periodogram of the raw and normalized PSM datasets, following band-pass filtering. In Fig. 2e, there is a low-frequency peak apparent at 0.34 Hz that overpowers the two peaks at 0.75 and 1.5 Hz that represent the RR at 45 bpm and its first harmonic. The effects of the Butterworth filter are visible by the roll-offs on both sides of the passband of 0.3-1.5 Hz and by the removal of the DC component in Fig. 2e. As shown in Fig. 2f, normalization cleans the lower frequency data, such that peak *a*, at the fundamental frequency, and peak *b*, at the second harmonic, are markedly visible. The fundamental frequency is then correctly recovered during frequency domain analysis.

The 20 trials were organized into three experimental conditions as shown in Table I. Each trial falls into one of the two classes within each condition. For example, in Trial 1 the simulator was lying supine on the overhead warmer mattress and breathing normally at 60 bpm. So, the results from Trial 1 contribute towards the aggregated results for supine, no-grunting, and overhead warmer. Table I lists the aggregate results of both time and frequency-domain analyses in terms of the RMS of RR-$FD_E$ and RR-$TD_E$. The RMS of RR-$FD_E$ is tabulated only for the raw dataset because, for each of three preprocessed datasets, the frequency domain analysis always estimated the correct RR producing an error of zero. Rather than reporting the zero error, Table I presents the RMS of $L_C$ for these three datasets. The values of $L_c$ provide a measure of estimation confidence (see Methods) and are informative when comparing experimental conditions in which the RR estimation errors are uniformly zero.

## IV. DISCUSSION

The pressure images obtained in this research using the simulator are comparable to those taken from real infants in [34]. The pressure image in Fig.1 shows regions of higher pressure in red which include the head, the thorax, and the pelvis. These results coincide with the regions of higher pressure shown in Fig. 9 of [34]. Breathing patterns in Fig. 2 are in agreement with the results obtained from real patient data in [22] in which Fig. 6 shows a normal or regular breathing pattern. The following observations can be made from Table I. The error in both time and frequency domains is highest across all the columns for the raw dataset as can be expected. With regards to position, the error for the raw dataset in both frequency and time domains is much higher for the supine position than for the prone position. This could be due to the location of the mechanical ventilator inside the simulator's body. Once the data are normalized, the results in the frequency domain for the supine position improve significantly; the error drops to zero with the RMS of $L_C$ being almost consistent across the three pre-processed datasets. The time domain results do not show a significant improvement in the supine position when the data are normalized and detrended. The highest improvement is seen when the data are median filtered. In the prone position, both domains perform well for normalized and detrended, however, the time domain worsens for the median filtered dataset whereas the frequency domain maintains a zero percent error and a steady Lc. In terms of breathing pattern, the RMS values of RR-$TD_E$ are much better for normal breathing as opposed to the noisy grunting pattern. Surprisingly, for the frequency domain approach, higher estimation confidence (Lc) is observed for breathing with grunting. In terms of mattress type, the firmer and thicker crib mattress results in better RR estimation across all three pre-processed datasets with lower RMS values of RR-$TD_E$ as compared to the overhead warmer mattress. Lc values remain steady across both mattresses in the pre-processed datasets indicating that the frequency domain approach is robust to mattress type.

One limitation of the present study is that it excluded an analysis of data in the presence of artifacts generated by patient motion and other electrical, mechanical, environmental and clinical variables. While this will be explored in future work, as shown here, pre-processing has played an important role in mitigating the impact of signal artifacts and improving the RR estimation rates for both domains.

This study produced results in the time domain that were comparable to the more complex compound time-frequency domain analyses developed in [16], [29]. The frequency domain results of this study are in agreement with past work done on adults in [10], and exceed the results produced in [16], [29]. In summary, the frequency domain analysis outperformed the time domain analysis in all three experimental conditions and across all neonatal RR evaluated here.

## V. CONCLUSION

This study compares time and frequency domain approaches to estimate RR from PSM data acquired during bench testing from neonatal patient simulators. The results clearly indicate that the frequency-domain approach is superior to the time domain approach. This research forms part of a larger novel project to assess the applicability of PSM technology in the NICU. In the future, we aim to assess RR estimation from real patient data. In addition, the fact that the estimation error is significantly higher during breathing with grunting may actually suggest a novel method for detecting grunting. A large difference observed between time- and frequency-domain RR estimates may indicate grunting. This can be further tested using model-based time-domain approaches that may show superior performance as compared to the time-domain methods applied in this paper. It is expected that sensor fusion between audio and PSM data may lead to a robust system for accurately identifying grunting during breathing, which is an important clinical indicator.

ACKNOWLEDGEMENT

The authors would like to acknowledge the assistance of CHEO NICU nurse Cheryl Aubertin during data collection.